\documentclass[11pt,preprint]{aastex}
\usepackage{epsfig}
\usepackage{color}

\begin{document}
\title {Particle Emission-Dependent Timing Noise of Pulsars?}

\author{\vspace{-\parskip}
LIU Xiong-Wei, NA Xue-Sen, XU Ren-Xin, QIAO Guo-Jun}
\affil{\vspace{-\parskip}
School of Physics and State Key Laboratory of Nuclear Physics and Technology,\\ Peking University, Beijing 100871, China;\\xiongwliu@163.com}

\begin{abstract}
Though pulsars spin regularly, the differences between the observed
and predicted ToA (time of arrival), known as ``timing noise'', can
still reach a few milliseconds or more. We try to understand the
noise in this study.
As proposed by Xu and Qiao in 2001, both dipole radiation and
particle emission would result in pulsar braking. Accordingly,
possible fluctuation of particle current flow is suggested here to
contribute significant ToA variation of pulsars. We find that the
particle emission fluctuation could lead to timing noise which cannot
be eliminated in timing process and that a longer period
fluctuation would arouse a stronger noise. The simulated timing
noise profile and amplitude are in agreement with the observed timing
behaviors on the timescale of years.
\end{abstract}

PACS:  97.60.Gb, 05.40.Ca

\vspace{0.5cm}
Why do pulsars spin down? This is a question still not fully answered even more than 40 years later since the discovery of the first pulsar.
It is generally suggested that pulsars spin down via magneto-dipole radiation, by which the ages and the surface magnetic fields are estimated accordingly.
However, it was proposed by Xu and Qiao (2001) that both dipole radiation and  relativistic particle emission powered by a unipolar generator can result in
the loss of pulsar rotation energy, and the observed braking indices ($<3$) could be understood then.$^{[1,2]}$ This opinion is consistent with later simulation$^{[3]}$
and observation.$^{[4]}$ In this Letter, we focus on further implication of the braking mechanism to timing behavior in Xu and Qiao's model.

Timing noise is the residual of pulsar time of arrival (ToA) after
fitted by the timing model. It reflects the effects of unknown elements
to ToA. A lot of models were proposed to explain timing noise, such
as the random walk in pulse frequency,$^{[5]}$ the free-precession
of neutron star,$^{[6]}$ the unmodelled companions,$^{[7,8]}$ and
the effect of gravitational waves.$^{[9]}$ However, the noise still
cannot be eliminated completely, especially in long timescale of
years.$^{[10-13]}$ On the other hand, the
pulsar flux density monitoring of the Green Bank$^{[14]}$ indicates
that the pulsar emission may not be absolutely stable. It is then
reasonable in Xu and Qiao's model that there exits fluctuation in the
relativistic particle emission, which would consequently contribute
to the timing noise. We will take the fluctuation in pulsar emission
into account in timing process, in this work, and try to find the
relationship between the fluctuation and timing noise.

The rotational energy loss rate is
\begin{equation}\label{eq:ZFebyM}
  -I\Omega\dot{\Omega}=\dot{E}=\dot{E_d}+\dot{E_u},
\end{equation}
where $I$ is the moment of inertia of a pulsar, $\Omega$ and $\dot{\Omega}$ are its angular velocity and the first derivative, $\dot{E}$ is the loss rate of rotational energy, and $\dot{E_d}$ and $\dot{E_u}$ are the powers of dipole radiation and relativistic particle flow, respectively.$^{[1]}$ When there is a fluctuation in $\dot{E_u}$, it becomes
\begin{equation}\label{eq:ZFebyM}
  \dot{E_u}=\bar{\dot{E_u}}(1+\delta),
\end{equation}
where $\bar{\dot{E_u}}$ is the stable value of $\dot{E_u}$, and $\delta$ is the fluctuation.
For different pulsars the relative quantities of $\dot{E_d}$ and $\dot{E_u}$ are different because the magnetic inclinations are distinct
and maybe the radiation mechanisms are not the same. However, for an individual pulsar these two components are sufficiently decided in a period of time and generally in a same order of magnitude. For the above reasons, and considering the dipole radiation is stable, we take
\begin{equation}\label{eq:ZFebyM}
  \dot{E_d}=n\times\bar{\dot{E_u}},
\end{equation}
where $n$ is a constant and decided by the magnetic inclination and radiation mechanism.

From Eqs. (1), (2) and (3) we obtain
\begin{equation}\label{eq:ZFebyM}
  -I\Omega\dot{\Omega}=\bar{\dot{E_u}}(n+1+\delta).
\end{equation}
Performing integration to both sides it becomes
\begin{equation}\label{eq:ZFebyM}
  {1\over2}I\left[{\Omega_0}^2-\Omega(T)^2\right]=\bar{\dot{E_u}}\left[(n+1)T+\int_{0} ^{T}\delta(t) dt\right],
\end{equation}
where $\Omega_0$ is the value of $\Omega$ at the beginning time, and we suppose that the moment of inertia $I$ is constant because it changes sufficiently small. When there is no fluctuation in $\dot{E_u}$, Eq. (5) becomes
\begin{equation}\label{eq:ZFebyM}
  {1\over2}I\left[{\Omega_0}^2-\Omega^{\prime}(T)^{2}\right]=\bar{\dot{E_u}}(n+1)T,
\end{equation}
where $\Omega^{\prime}(T)$ is the expected value when fluctuation is zero. Equation (5) minus Eq. (6) is
\begin{equation}\label{eq:ZFebyM}
  {1\over2}I\left[\Omega^{\prime}(T)^{2}-\Omega(T)^2\right]=\bar{\dot{E_u}}\int_{0} ^{T}\delta(t)dt.
\end{equation}
Considering the spin of pulsar changes very slowly, we obtain
\begin{equation}\label{eq:ZFebyM}
  \Omega^{\prime}(T)-\Omega(T)={\bar{\dot{E_u}} \over I\Omega_0} \int_{0} ^{T}\delta(t) dt.
\end{equation}
From Eqs. (4) and (8) we have
\begin{equation}\label{eq:ZFebyM}
  \Omega^{\prime}(T)-\Omega(T)={{-\dot\Omega_0} \over {n+1+\delta_0}} \int_{0} ^{T}\delta(t) dt.
\end{equation}
Performing integration to both sides we obtain
\begin{equation}\label{eq:ZFebyM}
{{-\dot\Omega_0} \over {n+1+\delta_0}}\int_{0} ^{\tau}\int_{0} ^{T}\delta(t) dtdT
= \int_{0} ^{\tau}\left[\Omega^{\prime}(T)-\Omega(T)\right]dT
= \Phi^{\prime}(\tau)-\Phi(\tau)
=-\Delta \Phi(\tau)
= -\Omega_0 R,
\end{equation}
where $\Phi$ is the phase of the pulsar, and $R$ is a provisional timing residual. Thus one has
\begin{equation}\label{eq:ZFebyM}
R={{\dot\Omega_0} \over {(n+1+\delta_0)}\Omega_0}\int_{0} ^{\tau}\int_{0} ^{T}\delta(t) dtdT
\\ =-{{\dot{P_0}} \over {(n+1+\delta_0)}P_0}\int_{0} ^{\tau}\int_{0} ^{T}\delta(t) dtdT,
\end{equation}
$P_0$ and $\dot{P_0}$ are the period and its first derivative at beginning time. Equation (11) reflects the relationship between the fluctuation and timing residual. We can obtain the real timing residual $\Re$ by performing least-squares-fitting to $R$.

To understand more clearly about Eq. (11), we try to provide a simple example. Let $\delta(t)=a\sin(2\pi t/t_0)$, one has
\begin{equation}\label{eq:ZFebyM}
\Re\cong{{a\dot{P_0}{t_0}^2} \over {4{\pi}^2(n+1+\delta_0)} P_0}\sin(2\pi {t\over t_0}).
\end{equation}
From Eq. (12) we can see that longer timescale variation will cause stronger noise because $\Re\propto {t_0}^2$. For a normal pulsar with $P_0=0.1$ s and $\dot{P_0}=1\times 10^{-14}$,
when $a=0.01$, $t_0=y\times3.15\times10^{7}$ s and $n=1$, we obtain $\Re \cong 0.013 \times y^2 \times \sin(2\pi t/t_0)$ s. It is a very strong noise at the timescale of years.

We further do a simulation with Eq. (11). Three sets of random
data with different Hurst parameter $H$, which reflects the time
dependence of a time series data,$^{[15]}$ are produced to simulate
three types of irregular fluctuations in $\dot{E_u}$. As is shown in
Fig. 1, each set of data has 10000 points. The first set has more
short period components, with $H=0.4$; the second set is approximate
white noise, with $H=0.6$; the third one has more long period
components, with $H=0.8$. In this simulation we take
$(n+1+\delta_0)P_0=0.1$ s and $\dot{P_0}=1\times 10^{-14}$. The
corresponding timing noises are shown in Fig. 2. The figures
indicate that if the particle emission has a random variation with
extent of about $1\%$ in daily timescale, the flux density from the
most distant pulsars varies less than $5\%$,$^{[14]}$ it will
lead to a timing noise with range of dozens of millisecond in 2000
days (shown on the left of Fig. 2), and several hundreds of millisecond in
10000 days (shown on the right of Fig. 2). These curves also show the
fluctuation with more long period components to cause stronger noise,
which accords with Eq. (12) very well.

Compared Fig. 2 with the observations, Fig. 1 in Ref. [11] and Figs. 1 and 2 in Ref. [12], we find that they have some common features. (1) The majority time curves have about one period-like
main structure no matter how long the time spans are (see Refs.\,[10,13]
for more examples), so that one cannot distinguish which one has the
long or short time span, just depend on their profiles, even for the
same pulsar. (2) The range from the minimum to maximum residual with longer time
span is larger than the one with shorter time span for each pulsar,
which is in agreement with Eq.\,(12). (3) The time curve of the shorter
time span is extremely similar to the corresponding
time span part of the longer one, which is natural because of the
integral relation in Eq.\,(11).

Recently, Lyne {\it et al}.$^{[16]}$ proposed another idea of producing timing
noise to pulsar, namely variations of the pulsar spin-down states variations lead
to timing noise. This phenomenologically explains the origin
of some quasi-periodic structures, which lie on lower-frequency
structures of some timing noise. However, it cannot give rise to the
ubiquitous lower-frequency structures in long time scales, which are
what we try to do in this study.

The statistics results from most pulsar timing noises are in agreement
with our model. Soon after we put our work on arXiv, a statistics
from Ryan {\it et al.} gives $\sigma
_{TN,2}\propto\nu^{0.9\pm0.2}{\mid\dot\nu\mid}^{1.0\pm0.05}
$,$^{[17]}$ which is consistent with Eq. (11) very well. From
observations, Cordes and Downs,$^{[18]}$ D'Alessandro {\it et
al.}$^{[10]}$ and Ryan {\it et al.}$^{[17]}$ all suggested that a mixture
of random walks in $\nu$ and $\dot{\nu}$ is compatible with the
timing noise, whereas we propose here a natural physical origin as
shown in Eq.\,(4). We can have the timing noises of millisecond
pulsar and AXP to be orders of $10^2$ ns and $10$ s, respectively,
from Eq. (11), which are consistent with the observations.

In summary, our model shows that the fluctuation of particle emission
will cause  significant timing noise. We emphasize that there could
be other kinds of the fluctuation (e.g., $\delta$), nevertheless the
long period composition of variation contributes larger to the
noise. The simulation accords with long (years) timescale noises both
in range and profile features. Simultaneously, our work supports the
opinion that the pulsar emission is not always stable, which is
important to the research of pulsar radiation and the understanding
of pulsar physics. Any other possible processes that lead to
instability to pulsar spin down energy could give timing residuals
similar to our result, and may be in agreement with the observations as
well as ours.

\acknowledgments
We thank the members at PKU pulsar group for helpful discussions. This work is supported by NSFC (10833003, 10935001, 10973002) and the National Basic Research Program of China under Grant No 2009CB824800.

\begin{figure}
\begin{center}
\epsfig{file=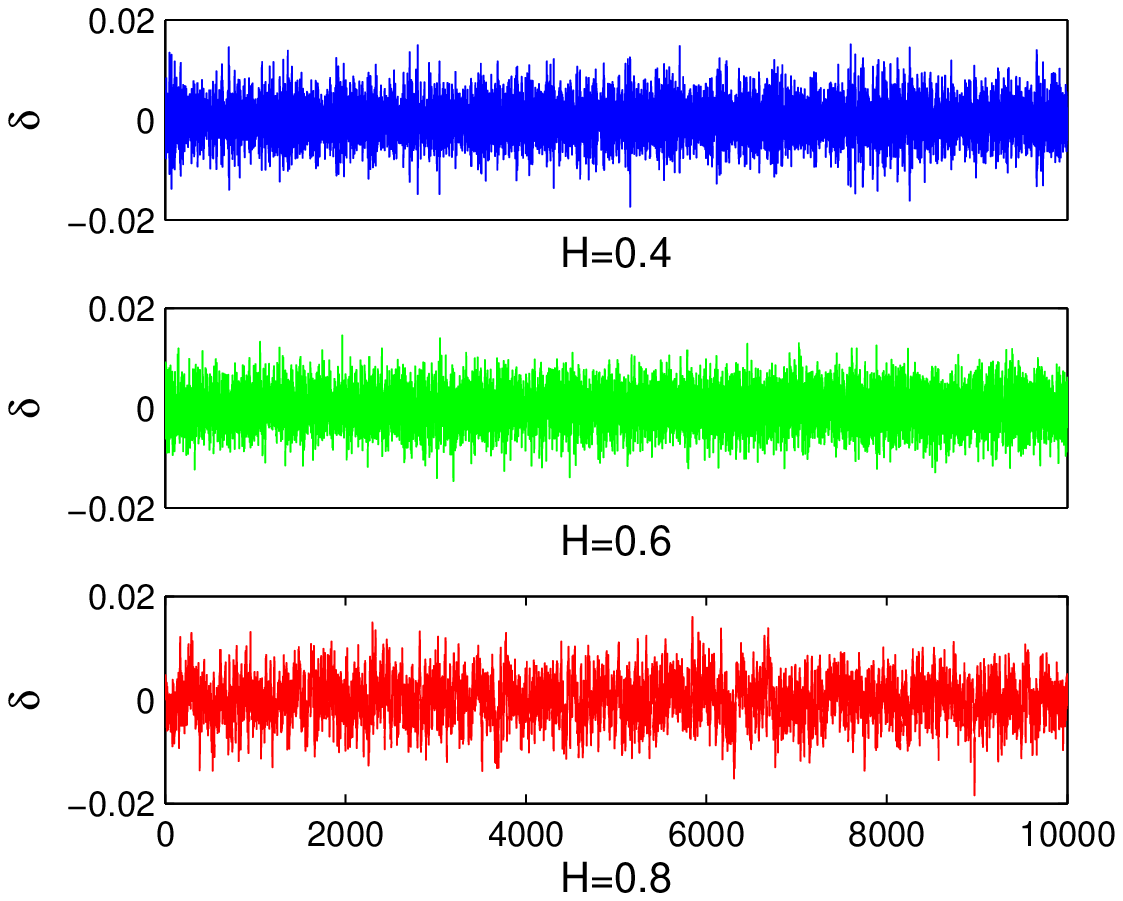,height=2.5in} \caption{The data used to simulate
the fluctuation of relativistic particles flux. The first set of
data has more short period component, the second set is approximate
white noise, the third set has more long period component. Here $H$
is the Hurst parameter.} \label{pres}
\end{center}
\end{figure}

\begin{figure}
\begin{center}
\epsfig{file=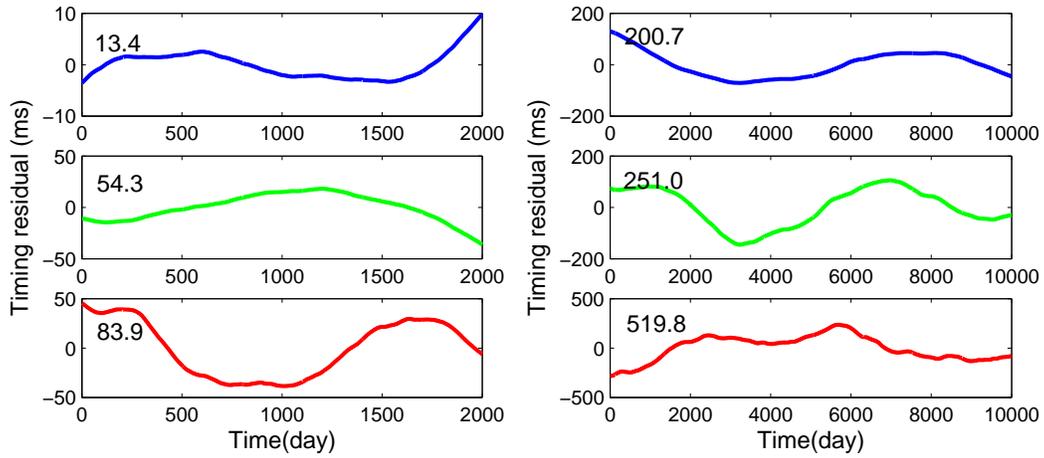,height=2.5in} \caption{Curves of timing
noise produced from the fluctuation data shown in Fig.\,1. The
Hurst parameters in the upper, middle, and bottom panels are $H=0.4,
0.6, 0.8$. The first 2000 points and the whole 10000 points are used
in the left and right panels, respectively. We take
$(n+1+\delta_0)P_0=0.1$\,s and $\dot{P_0}=1\times 10^{-14}$. The time
spans are about 5.5 and 27\,yr, and the label on the left of each
panel provides the range from the minimum to the maximum residual (ms).
The features of curve profile and noise rang are consistent with the
observations.} \label{pres}
\end{center}
\end{figure}

\end{document}